\documentclass{article}%
\usepackage{amsmath}
\usepackage{graphicx}%
\usepackage{amsfonts}%
\usepackage{amssymb}

\begin{document}

\title{A balanced gated-mode photon detector for qubit discrimination in 1550 nm}
\author{Akihisa Tomita\\Quantum Computation and Information Project, ERATO, JST
\and Kazuo Nakamura\\Fundamental Research Laboratories, NEC Corporation\\34 Miyukigaoka, Tsukuba, Ibaraki 305-8501, Japan}
\maketitle

\begin{abstract}
A photon detector combining the two avalanche photon diodes (APD) has been
demonstrated for qubit discrimination in 1550 nm. Spikes accompanied with the
signals in gated-mode were canceled by balanced output from the two APDs. The
spike cancellation enabled one to reduce the threshold in the discriminators,
and thus the gate pulse voltage. The dark count probability and afterpulse
probability were reduced to $7\times10^{-7}$ and $10^{-4}$, respectively,
without affecting the detection efficiency (11 \%) at 178 K.

\end{abstract}

One of the key devices for the optical implementation of quantum information
technology is a photon detector to determine the quantum states, or to
discriminate qubits. In discriminating qubits, we often split the photons
according to the state ($\left|  0\right\rangle $ or $\left|  1\right\rangle
$) and detect them. For example, qubits encoded on polarization states can be
discriminated by a polarization beam splitter followed by two photon
detectors. Phase information will be also obtained by detecting photons at the
two output ports of a Mach-Zehnder interferometer. We can find application of
the qubit discrimination in quantum key distribution (QKD)
experiments\cite{Marand95,Zbinden98,Bethune00,Hughes00}. The use of two
identical photon detectors is necessary for such demonstrations. Though a
single photon detector can act as two detectors by the use of time division
multiplexing technique\cite{Marand95}, this will increase the loss in the
detectors. Once we admit the requirement of two detectors, we should make the
most use of them. We show, in the following, the improved detector performance
by taking the differential signals of the two photon detectors.

The photon detectors should show high detection efficiency, low dark counts,
and short response time. The ratio of the detection efficiency $\eta$ to the
dark count probability $P_{d}$ determines the error rate, and so the range of
QKD transmission\cite{Zbinden98,Hiskett01,Stuck01,Bourennane01}. The ratio
$P_{d}/\eta$should be less than 10$^{-3}$ for 100 km fiber transmission in
1550 nm even with an ideal single photon source. Clock frequency is limited
mainly by the afterpulse, false photon detection caused by residual electrons
created by the previous detection. The QKD
experiments\cite{Marand95,Zbinden98,Bethune00,Hughes00} in 1550 nm have
employed the photon detectors using InGaAs/InP avalanche photodiodes (APDs) in
Gaiger mode\cite{Ribordy98}, where the reverse bias higher than the break down
voltage is applied. The high bias increases the avalanche gain to enable
single photon detection. However, this also results in large dark count
probability and afterpulse, which cause errors in the qubit discrimination.
The dark count probability and the afterpulse can be reduced by using
gated-mode\cite{Hiskett01,Stuck01,Bourennane01,Ribordy98}, where gate pulses
combined with DC bias are applied to the APD. The reverse bias exceeds the
break down voltage only in the short pulse duration. Though this method works
well, the short pulses produce strong spikes on the transient signals. High
threshold in the discriminator is therefore necessary to avoid errors, at the
cost of detection efficiency. High gate pulse voltage is also required to
obtain large signal amplitude by increasing avalanche gain. Impedance matching
helps to reduce the spikes to some extent\cite{Yoshizawa01}. Bethune and
Risk\cite{Bethune00} have introduced a coaxial cable reflection line to cancel
the spikes. We propose a much simpler method: canceling the spikes by taking
the balanced output of the two APDs required for the qubit discrimination.%

\begin{figure}
[h]
\begin{center}
\fbox{\includegraphics[
trim=0.000000in 0.000000in -0.002278in 0.000000in,
height=6.0978cm,
width=8.8019cm
]%
{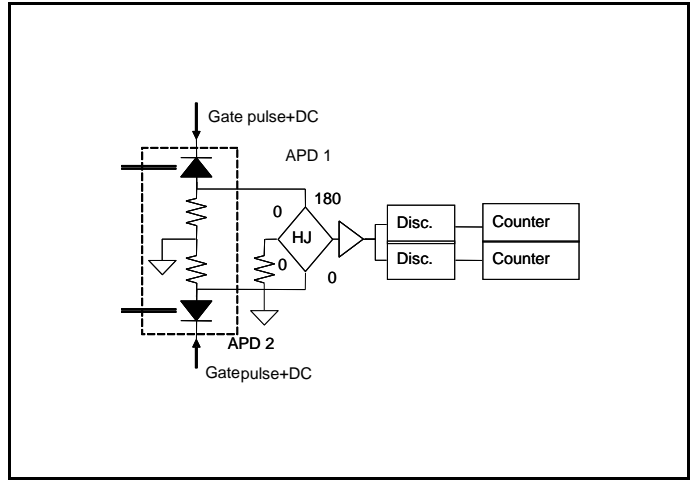}%
}\caption{Schematic of the photon detector. HJ and DISC stand for a hybrid
junction, and discriminators. The values of the resisters were 51 $\Omega$.}%
\label{schematic}%
\end{center}
\end{figure}

\begin{figure}
[h]
\begin{center}
\fbox{\includegraphics[
height=6.1cm,
width=8.8019cm
]%
{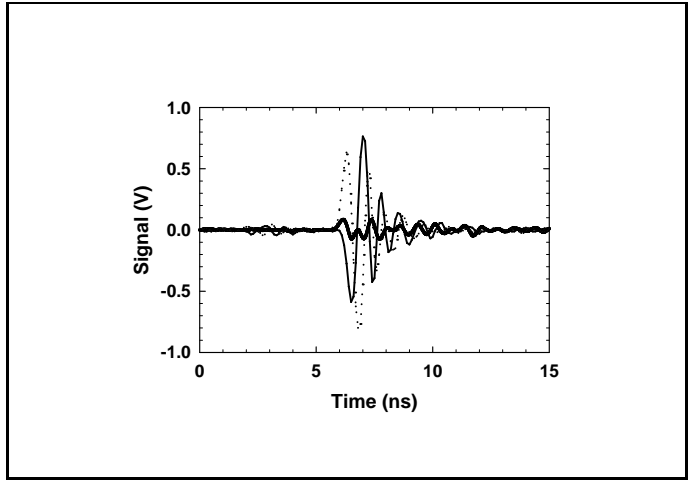}%
}\caption{Cancellation of the transient spike. Thin solid: APD 1, Dots: APD 2,
Thick solid: the differential output of the APD 1 and the APD 2.}%
\label{spike}%
\end{center}
\end{figure}

Figure \ref{schematic} depicts the schematic of the photon detector. Two APDs
(Epitaxx EPM239BA) and load resisters were cooled to 140 K-213 K by an
electric refrigerator. Short gate pulses of 2.5 V p-p and 750 ps duration were
applied to the APDs after being combined with DC bias by Bias-Tees. The output
signals from the APDs were subtracted by a 180${{}^{\circ}}$ hybrid junction
of 2-2000 MHz bandwidth (M/ACOM H-9.) The differential signal was amplified
and discriminated by two discriminators (ParkinElmer 9307.) Since the spikes
were the common mode input for the 180${{}^{\circ}}$ hybrid junction, they
would not appear at the output. The APD 1 provided negative signal pulses at
the output, while the APD 2 provided positive pulses. We can determine which
APD detects a photon from the sign of the output signals. Figure \ref{spike}
shows the output signal of the amplifier without photon input. Almost
identical I-V characteristics of the APDs enabled us to obtain a good
suppression of the spikes.%

\begin{figure}
[h]
\begin{center}
\fbox{\includegraphics[
height=6.1cm,
width=8.8019cm
]%
{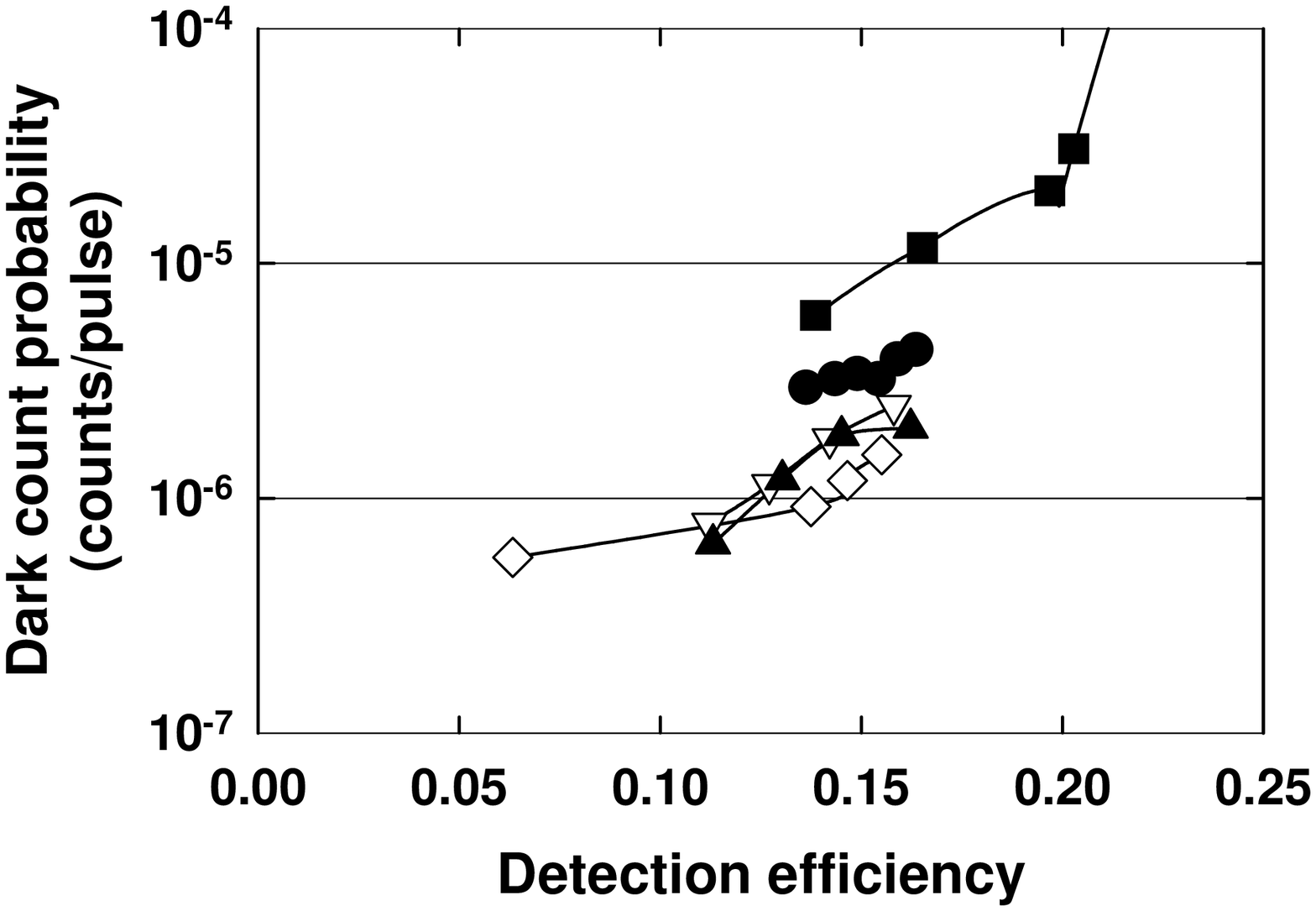}%
}\caption{Detection efficiency and dark count probability. Closed circles: APD
1 (140K), Open triangles: APD 1 (178 K), Squares: APD 1 (213 K), Diamonds: APD
2 (160 K), Closed triangles: APD 2 (178 K).}%
\label{qevsdark}%
\end{center}
\end{figure}

We first characterized the APDs separately. The attenuated light from a 1550
nm DFB laser was used as a photon source. The laser emitted the light pulse of
100 ps duration at the frequency of 100 kHz. The detection efficiency was
estimated by fitting to the theory assuming Poisson distribution of the
incident photons\cite{Levine84}. We accumulated pulse counts up to 10$^{8}$
clocks to obtain accurate dark count probability to the order of 10$^{-7}$. We
focused on the DC bias region where the detection efficiency larger than 10
\%. As shown in Fig. \ref{qevsdark}, we observed the lowest dark count
probability of 7x10$^{-7}$ per pulse with detection efficiency of 11 \% at 178
K. The ratio $P_{d}/\eta$ was as small as 6x10$^{-6}$, which corresponds to
220 km QKD transmission with an ideal photon source. To our knowledge, this is
the highest value of$P_{d}/\eta$. The detection efficiency and the dark count
probability are increasing functions of the bias. The maximum value of the
detection efficiency is obtained when the DC bias is set to the break down
voltage. We obtained larger values of the maximum detection efficiency at
higher temperatures: the detection efficiency of 20 \% at 213 K with the dark
count probability of 3x10$^{-5}$ per pulse.%

\begin{figure}
[h]
\begin{center}
\fbox{\includegraphics[
height=6.1cm,
width=8.8019cm
]%
{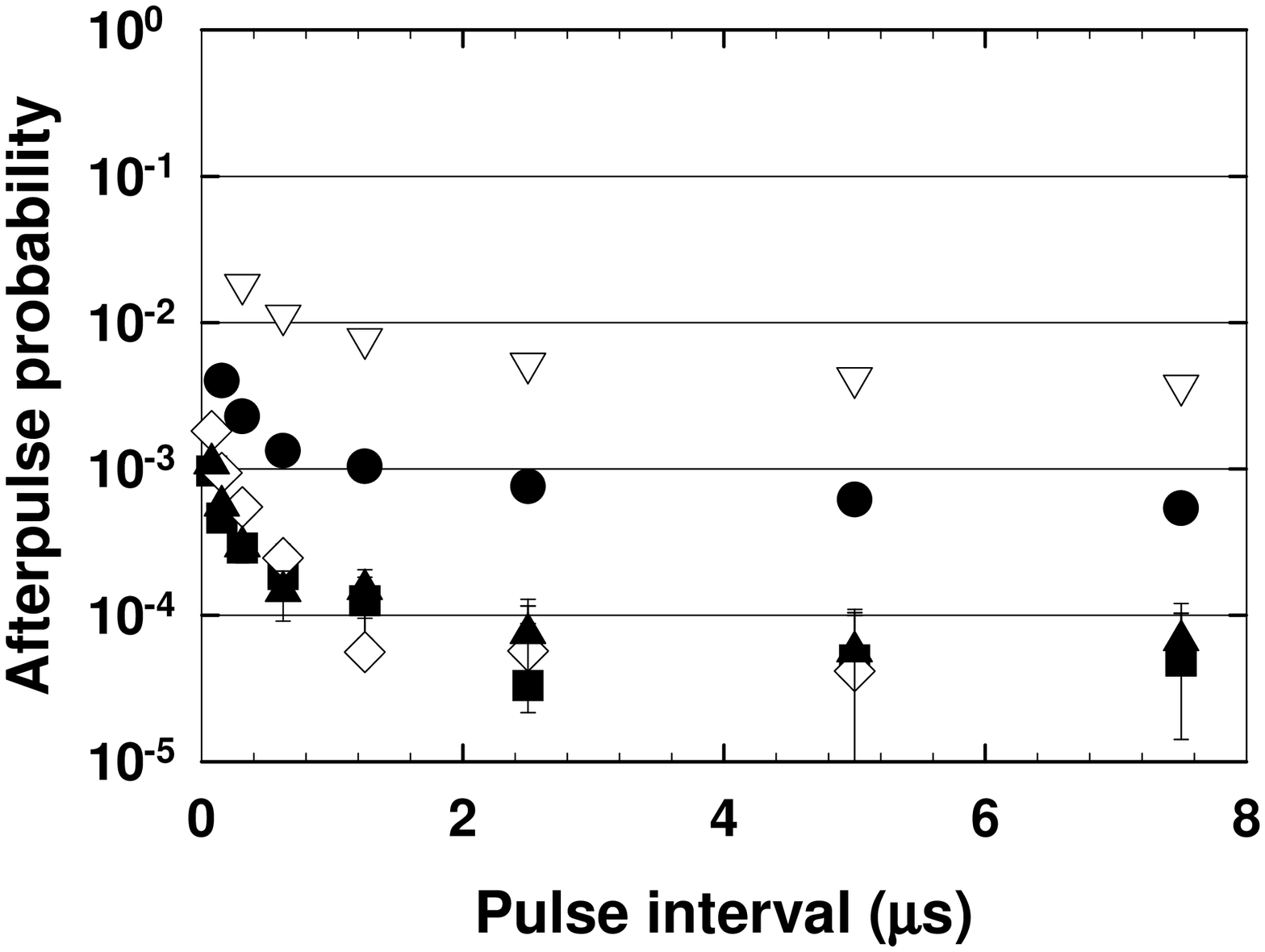}%
}\caption{Pulse interval and afterpulse probability. Closed circles: 160 K,
$\eta$=12 \%, Open triangles: 160 K, $\eta$=14 \%, Squares: 178 K, $\eta$=11
\%, Diamonds: 178 K, $\eta$=13 \%, Closed triangles: 213 K, $\eta$=14 \%.}%
\label{afterpulse}%
\end{center}
\end{figure}

Afterpulse probability was measured by applying two successive gate pulses to
the APDs. As seen in Fig. \ref{afterpulse}, afterpulse is prominent at low
temperatures. We found that afterpulse probability remained about 10$^{-4}$
for the 1 $\mu$s pulse interval at the temperatures higher than 178 K. This
corresponds to 10$^{-5}$ error probability (per pulse) for 10 \% detection
efficiency. Though Fig. \ref{afterpulse} shows the results on APD 1, we
obtained almost the same afterpulse characteristics on APD 2. Based on the
observation on the dark count probability and the afterpulse probability, we
conclude that the optimal operation temperature for the present APDs is around
178 K.

\begin{figure}
\begin{center}
\fbox{\includegraphics[
trim=0.000000in 0.000000in -0.002278in 0.000000in,
height=6.0978cm,
width=8.8019cm
]%
{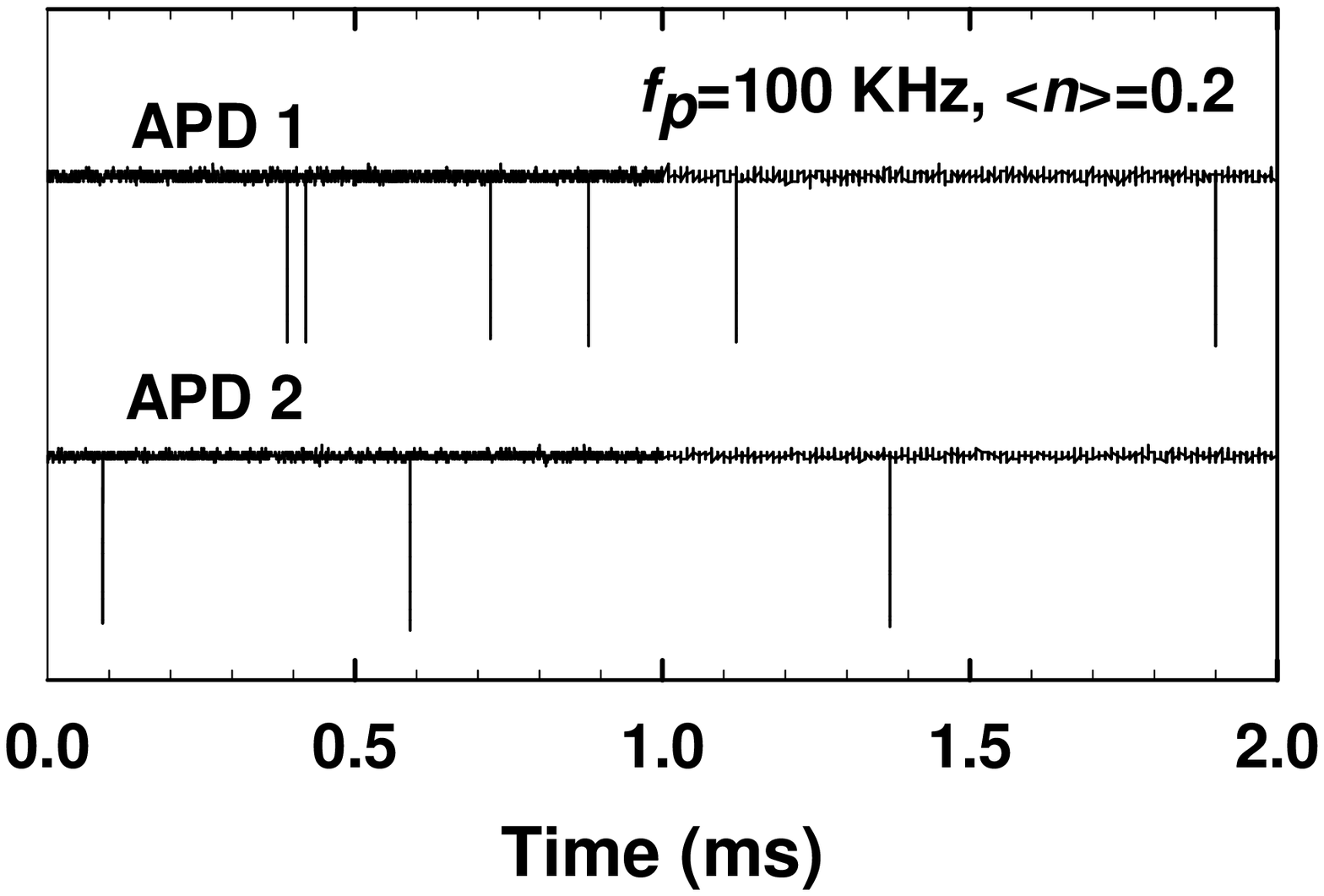}%
}
\caption{Photon detection in a `which path' experiment.}%
\label{whichpath}%
\end{center}
\end{figure}

Afterpulse also depends on the bias; the more afterpulse effect was observed
for the higher bias. This can be interpreted that higher bias produces more
electrons to contribute to the afterpulse during the avalanche process. The
present photon detector has shown a good afterpulse characteristic, because a
small bias voltage was enough to respond to a single photon. This is an effect
of the spike cancellation, by which we could reduce the threshold voltage in
the discriminators. We also found that afterpulse can be reduced by increasing
the DC bias with a constant total (DC + gate pulse) bias. The high DC bias may
sweep out the trapped electrons during the pulse interval.%

We observed photon detection by two APDs. The light was divided by a 3 dB
fiber coupler, and fed to the input ports of the APDs. In the photon picture,
the photon wave function will be collapsed by the photon detection at either
of the two APDs. This can be regarded as a `` which path'' experiment or the
latter part of a Mach-Zehnder interferometer. Figure \ref{whichpath} shows the
photon detection for 2 ms. Photon flux was set to 0.2 photons per pulse. APDs
were cooled to 178 K and the DC bias voltages were set to yield the detection
efficiency of 11 \%. We found that the detection probability of the two APDs
were almost the same for a weak light input. However, detection on the APD 1
became dominant as increasing photon flux. In the present experiment, APD 1
showed slightly larger detection efficiency than APD 2; so that APD 1 detected
predominantly the event that more than one photon entered the 3 dB coupler.
This observation implies that we should carefully adjust the detection
efficiency of the two APDs when more than one photon will be expected to be
detected. This is not the case in the qubit discrimination, however.

In conclusion, we have shown an improved photon detector for qubit
discrimination by taking the balanced output of two APDs. Canceling the spikes
enables to reduce the threshold at the discriminators, and thus decrease the
bias voltage. This reduces the dark count probability and the afterpulse
without sacrificing the detection efficiency. The present photon detector
would help to construct practical quantum information systems, such as QKD in
the lightwave communication wavelength.

\end{document}